\documentclass[10pt,conference]{IEEEtran}
\IEEEoverridecommandlockouts
\usepackage{cite}
\usepackage{amsmath,amssymb,amsfonts}
\usepackage{algorithmic}
\usepackage{graphicx}
\usepackage{textcomp}
\usepackage{xcolor}
\usepackage{listings}
\usepackage[most]{tcolorbox}
\usepackage{ragged2e}
\usepackage{url}
\usepackage{hyperref}
\usepackage{listings}
\usepackage{minted}

\lstdefinelanguage{json}{
    basicstyle=\ttfamily\footnotesize,
    numbers=left,
    numberstyle=\tiny,
    stepnumber=1,
    numbersep=5pt,
    showstringspaces=false,
    breaklines=true,
    frame=lines,
    backgroundcolor=\color{white},
    keywordstyle=\color{blue},
    stringstyle=\color{brown},
    morestring=[b]",
    literate=
     *{0}{{{\color{blue}0}}}{1}
      {1}{{{\color{blue}1}}}{1}
      {2}{{{\color{blue}2}}}{1}
      {3}{{{\color{blue}3}}}{1}
      {4}{{{\color{blue}4}}}{1}
      {5}{{{\color{blue}5}}}{1}
      {6}{{{\color{blue}6}}}{1}
      {7}{{{\color{blue}7}}}{1}
      {8}{{{\color{blue}8}}}{1}
      {9}{{{\color{blue}9}}}{1}
}

\def\BibTeX{{\rm B\kern-.05em{\sc i\kern-.025em b}\kern-.08em
    T\kern-.1667em\lower.7ex\hbox{E}\kern-.125emX}}

\begin{document}

\title{	SynRAG: A Large Language Model Framework for Executable Query Generation in Heterogeneous SIEM Systems\\


}

\makeatletter 
\newcommand{\linebreakand}{%
  \end{@IEEEauthorhalign}
  \hfill\mbox{}\par
  \mbox{}\hfill\begin{@IEEEauthorhalign}
}
\makeatother 




\author{
\IEEEauthorblockN{Md Hasan Saju\IEEEauthorrefmark{1},
Austin Page\IEEEauthorrefmark{2},
Akramul Azim\IEEEauthorrefmark{1},
Jeff Gardiner\IEEEauthorrefmark{2},
Farzaneh Abazari\IEEEauthorrefmark{2},
and Frank Eargle\IEEEauthorrefmark{2}}
\IEEEauthorblockA{\IEEEauthorrefmark{1}Department of Electrical, Computer, and Software Engineering (ECSE)\\
Ontario Tech University, Oshawa, Canada\\
\{mdhasan.saju, akramul.azim\}@ontariotechu.ca}
\IEEEauthorblockA{\IEEEauthorrefmark{2}GlassHouse Systems Inc., Toronto, Canada\\
\{apage, jgardiner, fabazari, feargle\}@ghsystems.com}
}


\maketitle

\begin{abstract}
Security Information and Event Management (SIEM) systems are essential for large enterprises to monitor their IT infrastructure by ingesting and analyzing millions of logs and events daily. Security Operations Center (SOC) analysts are tasked with monitoring and analyzing this vast data to identify potential threats and take preventive actions to protect enterprise assets. However, the diversity among SIEM platforms, such as Palo Alto Networks Qradar, Google SecOps, Splunk, Microsoft Sentinel and the Elastic Stack, poses significant challenges. As these systems differ in attributes, architecture, and query languages, making it difficult for analysts to effectively monitor multiple platforms without undergoing extensive training or forcing enterprises to expand their workforce.

To address this issue, we introduce SynRAG, a unified framework that automatically generates threat detection or incident investigation queries for multiple SIEM platforms from a platform-agnostic specification. SynRAG can generate platform-specific queries from a single high-level specification written by analysts. Without SynRAG, analysts would need to manually write separate queries for each SIEM platform, since query languages vary significantly across systems. This framework enables seamless threat detection and incident investigation across heterogeneous SIEM environments, reducing the need for specialized training and manual query translation. We evaluate SynRAG against state-of-the-art language models, including GPT, Llama, DeepSeek, Gemma, and Claude, using Qradar and SecOps as representative SIEM systems. Our results demonstrate that SynRAG generates significantly better queries for cross-SIEM threat detection and incident investigation compared to the state-of-the-art base models.

\end{abstract}


\section{Introduction}
 \label{section:introduction}
Cyberattacks have become increasingly common in recent years due to the rapid growth in internet usage and the widespread migration of sensitive and high-value systems to online platforms. A successful cyberattack can lead to millions of dollars in losses, damage to reputation, and erosion of client trust for an enterprise. As a result, organisations across industries have adopted Security Information and Event Management (SIEM) systems to proactively monitor and protect their digital infrastructure. SIEM systems collect, parse, and store large volumes of event and log data generated by an enterprise’s IT environment. Security analysts then examine this data to detect indicators of compromise, investigate potential threats, incidents, and respond to them in a timely manner. However, this task is becoming increasingly challenging due to the sheer volume of data and the diversity of SIEM platforms. Popular SIEMs such as QRadar from Palo Alto Networks (Previously owned by IBM), Google SecOps, Splunk, Apache Metron, and Microsoft Sentinel each have their own unique architectures, data models, interfaces, and query languages. Consequently, analysts must undergo significant training to become proficient in each individual SIEM system—a process that is both time-consuming and resource-intensive.

One of our ongoing collaborations is with a cybersecurity company GlassHouse Systems Inc., which provides security services to multiple clients across various industries. In order to accommodate their clients’ diverse technology stacks, the company’s analysts must regularly operate across multiple SIEM platforms. Their primary responsibility is to review log and offense data within these systems to identify suspicious activity. They often need to write queries in the respective platform's language to investigate an offense or incident. However, it is highly impractical for a single analyst to maintain deep expertise across all SIEMs simultaneously. They often need help delegating parts of the task to obtain queries for different SIEMs.

To address this challenge, we propose a novel framework called SynRAG, designed to streamline the generation of cross-SIEM threat detection queries. SynRAG allows security analysts to define potential threat scenarios using a platform-agnostic, structured YAML specification. From this unified input, SynRAG automatically generates platform-specific, executable queries for each supported SIEM system. These queries can be executed within their respective environments, with the results returned to the analyst in a standardized format. This approach not only reduces the need for specialized training and manual query rewriting, but also enhances threat detection efficiency across heterogeneous SIEM infrastructures by enabling centralized access to threat data.

Our key contributions are as follows:

\begin{itemize}
\item We introduce SynRAG, a novel framework that automatically generates executable, platform-specific SIEM queries from high-level, platform-independent threat specifications written in YAML.
\item To the best of our knowledge, SynRAG is the first research work designed to assist security analysts in investigating security incidents by generating valid queries tailored to different SIEM platforms.
\item Establishing a benchmark to support future research in automated SIEM query generation.
\end{itemize}

\section{Background and Related Works}
\label{section:related}
\textbf{Security Information and Event Management (SIEM):} SIEM is a centralized system that collects and analyzes log and event data from an organization’s IT infrastructure. Using that data it enables real-time threat detection, incident response, and compliance reporting. Modern SIEM platforms incorporate AI and machine learning to identify anomalies and automate security workflows. SIEM plays a critical role in safeguarding security operations centers (SOCs), especially in complex, hybrid IT environments.

\textbf{Logs, Events and SIEM Query:} Logs are timestamped records and activities generated by systems, applications, and devices that capture operational and security-related activities.
Events are significant actions or occurrences, such as logins, file access, or errors which are extracted from logs and analyzed for security relevance. Events are generated from the log data.
A SIEM query is a structured command used to retrieve, filter, and correlate event data from the SIEM platform to detect threats or investigate incidents. Millions of log and event records are ingested by SIEM systems every hour, making manual review impossible. Each SIEM platform provides its own query language to analyze events and detect threats or anomalies. Security analysts craft queries tailored to specific attack patterns or investigative needs for the safeguarding of the infrastructure.

\textbf{Retrieval Augmented Generation (RAG):} 
RAG\cite{lewis2020retrieval} is a technique that enhances language model outputs by first retrieving relevant external knowledge and then using it as context during text generation. In the context of SIEM query generation, this approach is crucial because each platform has strict, unique syntax rules and domain-specific tokens. The LLM models are not specially trained to generate SIEM queries. That's why instead of relying solely on the model’s internal knowledge, retrieving accurate documentation or syntax references ensures that the generated queries are both valid and executable. 

Several prior efforts have addressed the challenge of translating structured or semi-structured input into executable queries, particularly in domains such as knowledge base question answering, information retrieval, and program synthesis.

Zafar et al.~\cite{zafar2018sqg} introduced SQG, a modular system for generating SPARQL queries from natural language questions over RDF-based knowledge graphs like DBpedia. Their approach uses Tree-LSTMs to rank candidate queries based on syntactic and semantic alignment, focusing on semantic web use cases. However, SQG is limited to SPARQL and does not accommodate diverse platform-specific query syntaxes. Xue and Croft~\cite{xue2009patent} proposed a high-recall query formulation technique for patent retrieval. Their method uses tf-idf-weighted noun phrases from selected patent fields and applies learning-to-rank strategies. While effective for recall in static document retrieval, it does not support executable or platform-specific query generation. Huang et al.~\cite{huang2018metalearning} developed a meta-learning-based SQL generation system, PT-MAML, which performs few-shot adaptation via pseudo-task creation based on question similarity. Though it also targets structured query generation, it relies on gradient updates and is less suited for real-time applications in security contexts. In contrast to these, SynRAG focuses on query generation to detect threat and investigate incidents across SIEM platforms from a YAML-based specs using a Retrieval-Augmented Generation (RAG) based pipeline. No other studies specifically worked on SIEM based query generation, so we tried to fill that gap in our work.

There has been very limited work focused on transforming high-level user requirements into executable SIEM queries that assist SOC analysts in investigating threats within their infrastructure. Our work sets the foundation for this capability by enabling the automatic generation of platform-specific SIEM queries from abstract, human-readable threat specifications.

\section{Methodology} \label{section:methodology}

\begin{figure*}[t]
\centering
\includegraphics[width=\textwidth, height=0.35\textheight, keepaspectratio]{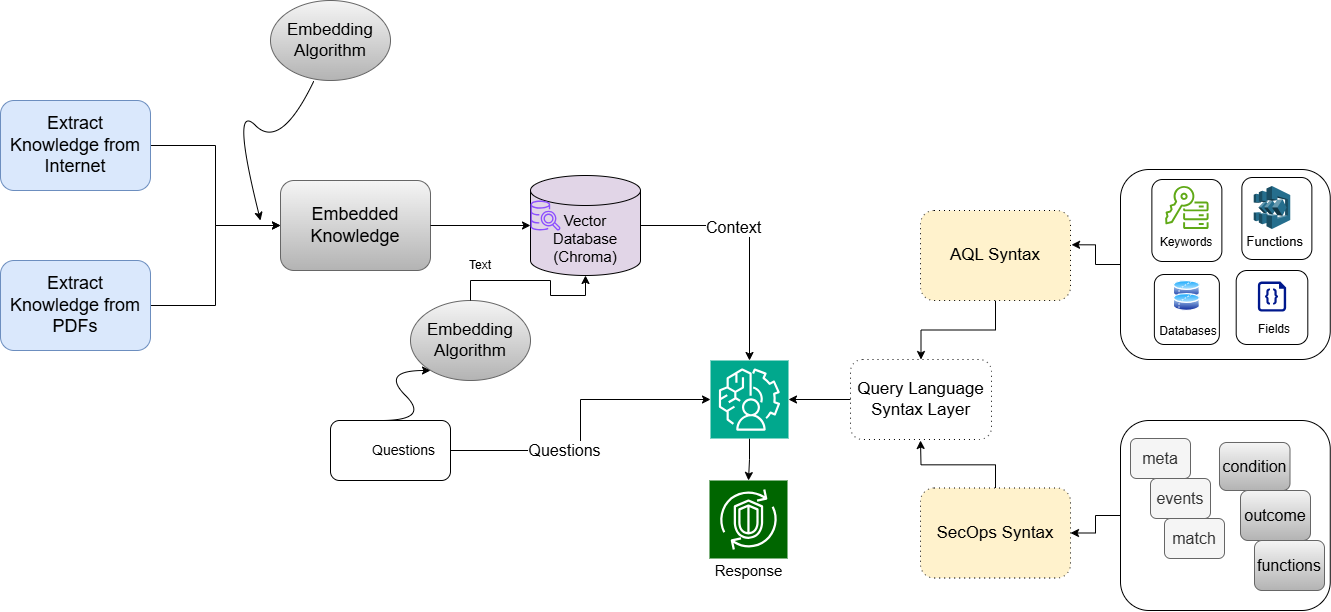}
\caption{SynRAG Architecture}
\label{fig:synRag}
\end{figure*}

In this section, we describe the methodology behind SynRAG(figure \ref{fig:synRag}) and how it enables the generation of platform-specific queries for various SIEM systems from a SIEM-agnostic specification.

SIEMs can ingest millions of logs and events per day, and while many platforms allow rules to be created for ongoing threat detection, defining and deploying rules takes time and effort. In contrast, running a one-time, on-demand query can quickly surface evidence of a specific threat. Each SIEM platform has its own unique query language. For example: Palo Alto Networks QRadar uses AQL (Ariel Query Language) which is previously owned by IBM, Google SecOps uses YARA-L 2.0, Splunk uses Search Processing Language (SPL), Elastic Stack uses ES|QL (Elasticsearch Query Language).
These syntactic and structural differences make it difficult for analysts to efficiently craft queries across multiple systems. 

Consider the example of a potential brute-force login attack:
A user attempts to log in 20 times within 5 minutes over the past 5 hours.
To investigate this behavior, a query can be written in the native language of the SIEM in use. However, with 7–8 different SIEM platforms in use across an enterprise, it becomes impossible for analysts to maintain proficiency in all query languages.

This is where SynRAG provides a significant advantage. SynRAG uses a yaml file as requirements for describing log-based detection logic in a platform-independent way which allows analysts to describe threats using a unified format. The file defines conditions such as event types, field values, temporal constraints, and other detection logic in plain English language without binding to a specific SIEM’s syntax or structure.

Our approach involves two main steps:

\label{inputDataset}
\textbf{Threat Specification:} Analysts define threat behavior in a YAML file using natural human language in a specific format which includes description, fields to select, source, logic. In collaboration with a human expert, we developed 40 threat specifications to evaluate our architecture, a process that demands substantial time and effort. 

\textbf{Query Generation:} SynRAG parses the rule and translates it into native queries for each supported SIEM platform.

In our current implementation, we focus on IBM QRadar and Google SecOps as representative SIEM systems. As this is an ongoing project, support for additional SIEM platforms will be added progressively over time. SynRAG can generate the corresponding AQL and YARA-L 2.0 queries from the specification. We can validate the correctness of the generated queries by executing them in their respective SIEM environments or by comparing the generated query with a standard query written by a human expert. By automating this translation process, SynRAG empowers analysts to detect threats or investigate incidents across different SIEM systems without needing to manually learn and write queries in each platform’s proprietary language.

\subsubsection{Architecture Overview}
The proposed system utilizes a Retrieval-Augmented Generation (RAG)\cite{lewis2020retrieval} architecture to automatically generate platform-specific security queries based on high-level threat descriptions. The methodology(figure \ref{fig:synRag}) comprises three major stages: Knowledge Extraction, Vector Database Creation, and Syntax Service.

First, domain knowledge regarding the query languages used by QRadar (AQL) and Google Chronicle SecOps (YARA-L) was collected. Next, the collected documents were embedded and stored in a persistent vector store using the Chroma database. To support syntax-aware generation, a syntax service was developed to provide curated lists of valid syntax and semantic constraints for each query language. At inference time, when a user provides a natural language specification of a threat scenario which is marked as questions in figure \ref{fig:synRag}, the input is embedded and used to perform semantic retrieval over the Chroma vector database. The system retrieves the top-5 relevant knowledge chunks for that specific platform (QRadar or SecOps). Simultaneously, the syntax service provides the allowed query components for the corresponding platform. The language model then conditions its output on both the retrieved context and platform-specific syntax. Using this combined input, it generates a syntactically valid and semantically aligned query for each platform independently. This process ensures that the generated queries are both compliant with platform requirements and aligned with the user’s intent.


\subsubsection{Knowledge Extraction}

To support the RAG (Retrieval-Augmented Generation) architecture, a domain-specific knowledge base was constructed from two primary sources: QRadar AQL documentation in PDF format and Google SecOps documentation hosted online. Two separate Python-based data collection pipelines were implemented to handle the format and structure of each source.

\textbf{QRadar AQL Knowledge Extraction (Offline PDF Sources):}
We extracted AQL documentation from PDF files using a custom Python script built with the \texttt{PyMuPDF} (\texttt{fitz}) library. The script recursively processed all \texttt{.pdf} files in a target directory, extracted text from each page, and saved the content as UTF-8 encoded \texttt{.md} (Markdown) files. This conversion preserved document boundaries and prepared the data for downstream parsing and indexing.

\textbf{Google SecOps Knowledge Extraction (Online HTML Sources):}
An asynchronous web scraping pipeline, built with \texttt{Playwright}, was used to extract documentation from Google SecOps’s portal. The crawler discovered valid English URLs, launched a headless Chromium browser to extract content within the \texttt{<article>} tag, and saved it as Markdown files. Filenames were derived from sanitized page titles and included source metadata. This standardized collection of documents was used to build the knowledge base for the RAG retriever, enabling efficient semantic retrieval.

\subsubsection{Vector Database Creation}

To enable semantic retrieval within the RAG pipeline, a dense vector index was constructed from the preprocessed Markdown documents. All Markdown files, previously extracted from both QRadar (PDF) and SecOps (web) sources, were loaded using the \texttt{UnstructuredMarkdownLoader} module from the \texttt{langchain\_community.document\_loaders} package. Each document was enriched with metadata, including the source path, origin folder, and a unique \texttt{UUID} identifier to ensure traceability. For effective indexing, documents were segmented using the \texttt{RecursiveCharacterTextSplitter}, with a chunk size of 500 characters and an overlap of 100 characters. This overlap helps retain contextual coherence across boundaries and improves retrieval performance. The resulting text chunks were embedded using \texttt{HuggingFaceEmbeddings}, configured with the pre-trained model \texttt{sentence-transformers/all-MiniLM-L6-v2}. These dense vectors, along with their corresponding metadata, were stored in a \texttt{Chroma} vector database. A detailed logging mechanism was implemented to record file loading durations, errors, and the total number of vectorized chunks, providing transparency and reproducibility for the data indexing pipeline.

\subsubsection{Syntax Service}
A common issue with generating SIEM queries using LLMs is hallucination. They often fabricate or invent syntax. To prevent this, we introduced our Syntax Service.

\textbf{AQL Syntax Segmentation:} The query language used by QRadar is the Ariel Query Language (AQL). To enable accurate and context-aware query generation, the AQL syntax was deconstructed into four fundamental components:

\begin{enumerate}
    \item \textbf{Keywords} – Reserved terms that define query actions (e.g., \texttt{SELECT}, \texttt{FROM}, \texttt{WHERE}).
    \item \textbf{Field Names} – Schema-specific attribute identifiers (e.g., \texttt{sourceIP}, \texttt{destinationPort}).
    \item \textbf{Functions} – Built-in operations for filtering and transformation (e.g., \texttt{UTF8}, \texttt{LOWER}, \texttt{CONTAINS}).
    \item \textbf{Database/Source} – The log source or table on which the query is executed (e.g., \texttt{events}, \texttt{offenses}).
\end{enumerate}

Each valid AQL query can be expressed as a composition of elements from these four categories. To enforce syntactic validity and domain conformity during generation, a constrained vocabulary set was curated for each component. These curated token sets were then embedded into the RAG pipeline, allowing the language model to condition its output solely on this predefined syntax space. By restricting the generative process to authorized AQL syntax elements, the system ensured that all model-generated queries adhered to QRadar's execution standards, reduced hallucination risk.

\textbf{YARA-L Query Syntax Segmentation:} YARA-L 2.0 is the query language used in Google Security Operations for analyzing structured event data based on the Unified Data Model (UDM). In this work, we decomposed YARA-L syntax into core components—\textbf{meta}, \textbf{events}, \textbf{match}, \textbf{condition}, \textbf{functions} (e.g., \texttt{re.capture()}), and \textbf{outcome}—to support syntax-constrained generation within a Retrieval-Augmented Generation (RAG) framework. We curated a structured token set for all the core components from official documentation and embedded it into the RAG pipeline. By enforcing adherence to YARA-L grammar rules, including correct use of placeholders and logical expressions, our approach ensured the generation of syntactically valid and executable queries. This significantly reduced errors and improved the reliability of the output queries.

\section{Experimental Evaluation}
\subsection{Experimental Setup}

All experiments were conducted on a personal machine equipped with a 12th Gen Intel\textsuperscript{\textregistered} Core\textsuperscript{TM} i5-1240P processor running at 1.70 GHz, 16 GB of installed RAM, and an NVIDIA GeForce MX550 GPU with 2 GB of dedicated memory and 7.9 GB of shared memory. The GPU was used to accelerate embedding and inference tasks.

We evaluated multiple large language models via their respective APIs to ensure fairness and consistency. The models include \texttt{GPT-4o}, \texttt{deepseek-ai/DeepSeek-V3}, \texttt{meta-llama/Llama-3.3-70B-Instruct-Turbo}, \texttt{google/gemma-2-27b-it}, and \texttt{Claude Sonnet 4}. SynRAG also used \texttt{GPT-4o} as the base model, integrated within a Retrieval-Augmented Generation (RAG) framework that provides task-specific context and syntax constraints. All model queries were executed through API endpoints.

\textbf{Prompt Engineering: } 
Prompt engineering\cite{white2023prompt} plays an important role in guiding large language models (LLMs) to generate accurate and context-aware outputs, especially in domain-specific tasks like security query generation. A well-crafted prompt can significantly reduce ambiguity, enforce syntax constraints, and improve the semantic alignment of the model's response with the intended goal.

To achieve more accurate results, writing precise prompts is crucial. For both AQL and SecOps, we designed two separate prompts. The AQL prompt is added in Figure~\ref{fig:aql_prompt}. These prompts were selected based on a trial-and-error approach, iteratively refined to produce syntactically valid and semantically correct queries within the constraints of each SIEM platform. All test prompts, the SecOp prompt, sample YAML detection logics, source code, and resources for generating the vector database are provided at the link included in the footnote\footnote{\url{https://github.com/Hasan-Saju/siem-independent-query}}.

\subsection{Evaluation Metrics}

We used two widely adopted metrics to evaluate the quality of the generated SIEM queries: \textbf{BLEU} and \textbf{ROUGE-L}\cite{yang2018adaptations}. The BLEU (Bilingual Evaluation Understudy) score measures the degree of n-gram overlap between the generated query and a reference query. It is sensitive to exact token matches and is commonly used to assess syntactic similarity in generation tasks. A higher BLEU score indicates that the output closely matches the structure of the reference. The reference was written by a subject matter expert with a strong understanding of that SIEM query language.

ROUGE-L, on the other hand, focuses on the longest common subsequence between the generated and reference queries. It is more tolerant to changes in word order and is used to measure the semantic and lexical similarity of the output. ROUGE-L is especially useful in our context because SIEM queries may still be valid even if their tokens appear in a different order, as long as their intent and structure are preserved. Together, these two metrics provide a balanced view of both syntactic correctness and semantic fidelity of the generated queries.

\subsection{Results and Discussion}

\begin{figure}[t]
\centering
\includegraphics[height=0.33\textheight]{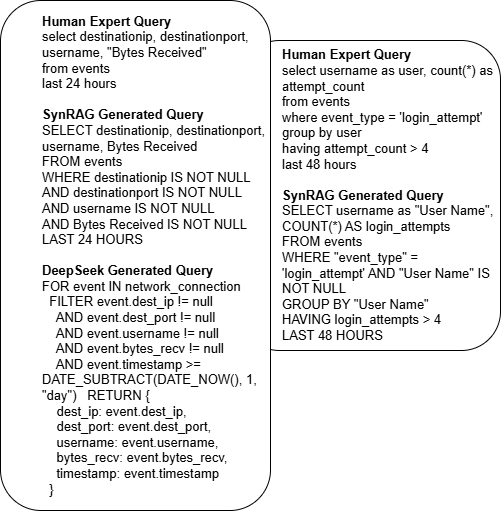}
\caption{Output Query Comparison}
\label{fig:resultVis}
\end{figure}

We evaluated our SynRAG framework and several strong baseline models on the task of cross-SIEM threat detection query generation using the specification set mentioned in the methodology,\ref{inputDataset}. Two metrics were used for evaluation: BLEU and ROUGE-L.


\begin{table*}[!htbp]
\centering
\caption{Performance Across Runs and Averages}
\begin{tabular}{|l|cc|cc|cc|cc|}
\hline
\textbf{Model} & \multicolumn{2}{c|}{\textbf{Run 1}} & \multicolumn{2}{c|}{\textbf{Run 2}} & \multicolumn{2}{c|}{\textbf{Run 3}} & \multicolumn{2}{c|}{\textbf{Average}} \\
 & \textbf{BLEU} & \textbf{ROUGE-L} & \textbf{BLEU} & \textbf{ROUGE-L} & \textbf{BLEU} & \textbf{ROUGE-L} & \textbf{BLEU} & \textbf{ROUGE-L} \\
\hline
GPT         & 0.0431 & 0.4184 & 0.0431 & 0.4226 & 0.0539 & 0.4439 & 0.0467 & 0.4283 \\
Claude      & 0.0272 & 0.2844 & 0.0275 & 0.2927 & 0.0365 & 0.3417 & 0.0304 & 0.3063 \\
LLaMA       & 0.0777 & 0.5048 & 0.0512 & 0.4881 & 0.0818 & 0.4587 & 0.0702 & 0.4839 \\
DeepSeek    & 0.0396 & 0.3565 & 0.0420 & 0.3892 & 0.0568 & 0.3923 & 0.0461 & 0.3793 \\
Gemma       & 0.0575 & 0.3170 & 0.0658 & 0.3845 & 0.0446 & 0.3759 & 0.0560 & 0.3591 \\
\textbf{SynRAG (Ours)} & \textbf{0.1243} & \textbf{0.6076} & \textbf{0.1008} & \textbf{0.5736} & \textbf{0.1610} & \textbf{0.6307} & \textbf{0.1287} & \textbf{0.6039} \\
\hline
\end{tabular}
\label{tab:model_performance}
\end{table*}

Table \ref{tab:model_performance} reports the performance of SynRAG and other language models across three independent runs using BLEU and ROUGE-L as evaluation metrics. Our proposed model, SynRAG, demonstrates clear and consistent superiority over all baselines. It achieves the highest average BLEU score of 0.1287 and ROUGE-L score of 0.6039, significantly outperforming the closest competitor, LLaMA, which attains 0.0702 BLEU and 0.4839 ROUGE-L on average. The standard models (GPT, Claude, DeepSeek, and Gemma) exhibit relatively lower and more variable results. SynRAG also exhibits low variance, indicating robustness and stability. We also executed the generated queries from each model and found that in \textbf{85\% of the cases, SynRAG’s queries executed successfully without any syntax issues}, while the remaining 15\% required only minimal changes (e.g., missing quotation marks or minor spelling adjustments). In contrast,\textbf{ all the queries generated by all other models required at least some modifications}—such as correcting field names, query structure, or function usage—before they could be executed. This shows that SynRAG significantly performs better. That means our architecture outperforms all other baseline models in generating SIEM queries for investigating security incidents.

We evaluated each model over three runs, rather than a single instance. It helps mitigate the effects of variability inherent in generation-based tasks, leading to a more reliable and statistically sound comparison. These results underscore the effectiveness of SynRAG in generating outputs with higher lexical and semantic fidelity. The performance gap between SynRAG and the other baseline models is mainly due to two key strengths of SynRAG:

\begin{itemize}
    \item \textbf{Domain-Aware RAG Architecture:} SynRAG leverages a Retrieval-Augmented Generation pipeline that uses actual SIEM documentation to provide context during query generation. This reduces hallucinations and improves factual correctness in query syntax.

    \item \textbf{Syntax-Constrained Generation:} SynRAG uses a curated syntax service to enforce correct use of query language tokens. This helps the model generate executable queries using only the provided keywords and syntax, which in turn reduces the likelihood of hallucinating new or incorrect syntax.

\end{itemize}


In the output comparison illustrated in Figure \ref{fig:resultVis}, SynRAG exhibits strong performance in generating accurate and executable SIEM queries. In the first scenario, SynRAG’s query mirrors the human expert’s intent by capturing the correct structure and logic. The only minor flaw lies in the omission of quotation marks around "Bytes Received", a syntactic requirement for multi-word field names in some platforms like QRadar. Interestingly, SynRAG also introduces additional IS NOT NULL conditions for each field, which, although not present in the expert version, improve query robustness by ensuring completeness of results. In contrast, the base model DeepSeek produces a query that is neither syntactically valid nor semantically aligned—it uses wrong field names and applies an entirely incorrect structure, rendering the output unusable in a real Qradar SIEM system.

In the second scenario of figure \ref{fig:resultVis}, focused on detecting brute-force login attempts, SynRAG generates a query that is functionally identical to the expert's version. While there are slight stylistic differences—such as alias formatting or column naming—the logic, filtering, aggregation, and grouping operations are all correctly implemented. This highlights SynRAG’s ability to produce not just valid queries, but ones that align with real-world analyst expectations. These results underscore the system’s effectiveness in practical security operations, where syntactic precision and semantic accuracy are both critical for actionable threat detection.

Overall, these results demonstrate that SynRAG is highly effective for the practical challenge of cross-SIEM query generation. Its strong performance in both syntactic and semantic metrics makes it a reliable tool for SOC analysts working with diverse security platforms.

\begin{figure}[t]
\centering
\begin{tcolorbox}[colback=white, colframe=black, title=AQL Query Generation Prompt, fonttitle=\bfseries, coltitle=white, boxsep=1mm, left=1mm, right=1mm, fontupper=\scriptsize]
\RaggedRight
\scriptsize

You are an expert in SIEM correlation and threat detection. Your task is to convert a detection description into a QRadar AQL query.

You will receive:
\begin{itemize}
    \item A Detection specification (YAML format)
    \item A list of allowed AQL fields, keywords, functions, and database names
    \item Context from other API call
\end{itemize}

Use the following strict instructions:

\begin{enumerate}
    \item Construct an AQL query using \textbf{only} the provided fields, keywords, functions, and databases.
    \item Use the correct AQL clause order: \texttt{SELECT} $\rightarrow$ \texttt{FROM} $\rightarrow$ \texttt{WHERE} $\rightarrow$ \texttt{LAST}
    
    \textbf{Note: Some instruction lines are omitted for space constraints.}


    \item Use only fields listed in the \texttt{fields} array. If a field from the detection rule isn't present, find the closest logical/semantic match or add a \texttt{-- TODO} comment.
\end{enumerate}

\textbf{Threat Detection Logic:} \\
\texttt{\$\{DetectionLogic from the YAML\}}

\textbf{Field List:} \\
\texttt{fields =  \{JSON.stringify(field list from API)\}}

\textbf{Keyword List:} \\
\texttt{keywords = \{JSON.stringify(keyword list from API)\}}

\textbf{Database List:} \\
\texttt{databases = \{JSON.stringify(database name from API)\}}

\textbf{Function List:} \\
\texttt{functions = \{JSON.stringify(function list from API)\}}

\textbf{Additional Context:} \\
\texttt{\$\{context from API || 'N/A'\}}

\textbf{Instruction:} \\
Generate the AQL query now.

\end{tcolorbox}
\caption{Instructional prompt for the AQL Query Generation task.}
\label{fig:aql_prompt}
\end{figure}

\section{Threat to Validity}
SynRAG currently supports only Palo Alto Network QRadar (previously owned by IBM) and Google SecOps. We still need to accommodate SynRAG for other popular platforms like Splunk, Elastic Stack, and Microsoft Sentinel. We are currently working on adding support for these, as this is an ongoing project. Again, our evaluation was conducted on a limited set of manually curated YAML-based detection rules, as creating these specifications requires significant time and effort. We are actively working on expanding the dataset and incorporating more comprehensive validation scenarios as part of our future work.


\section{Conclusion} \label{section:conclusions}
In this work, we presented SynRAG, a platform-agnostic framework for generating SIEM-specific threat detection queries from unified, high-level specifications. By leveraging a Retrieval-Augmented Generation (RAG) architecture combined with a query syntax constraint layer, SynRAG addresses the syntactic diversity and complexity of modern SIEM platforms. Our evaluation demonstrates that SynRAG significantly outperforms state-of-the-art large language models in both syntactic fidelity and semantic alignment, generating executable queries with high accuracy and robustness.

By reducing the need for manual query translation and platform-specific expertise, SynRAG empowers Security Operations Center (SOC) analysts to operate more efficiently across heterogeneous SIEM environments. This enables organizations to adopt a more unified and scalable approach to threat detection and incident investigation. As we expand support for additional SIEM platforms and broaden our benchmark dataset, SynRAG is poised to become a foundational tool for cross-SIEM security automation in modern enterprise settings.

\vspace{12pt}
\color{red}

\end{document}